\documentclass[11pt,twoside]{article}

\usepackage{asp2006}
\usepackage{epsf}
\usepackage{psfig}
\usepackage{lscape}
\usepackage{graphicx}

\begin{document}
\title{Variation of Oscillation Mode Parameters over the Solar Cycle 23: An
Analysis on Different Time Scales}
\author{S. C. Tripathy, F. Hill, K. Jain, and J. W. Leibacher}   
\affil{National Solar Observatory,  950 N. Cherry Avenue, Tucson, AZ 85719,
USA}   

\begin{abstract} 
We investigate the variation in the mode parameters obtained from time series
of length 9, 36, 72 and 108 days to understand the changes occurring on
different time-scales. The regression analysis between frequency shifts and 
activity proxies indicates that the correlation and slopes are correlated and
both increase in going from time series of 9 to 108 days. We also observe that
the energy of the mode
is anti-correlated with solar activity while the rate at which the energy is 
supplied remains constant over the solar cycle.    
\end{abstract}

\section{Introduction}  
Observations show that the mode frequencies change with the solar cycle and the
correlation  between the frequencies and solar activity measured by different
proxies is solar cycle phase dependent \citep[see][and references
therein]{jain09}. There
is  also an indication that the correlation between frequency and activity
depends on the length of the observation due to the finite lifetime of the
modes \citep{sct07}. Therefore, it is important to understand the relation
between mode frequencies and other mode parameters with solar activity on
different time scales.  With this objective, we have
analyzed the Global Oscillation Network Group (GONG) time series data by
splitting it into segments of length 9, 36, 72 and 108 days. 

\section{Data}
The GONG data analyzed here cover the period between 7 May 1995 and 31 August 
2007 and consist of 125 sets of time series 
of length 36 days. These are processed with a multi-taper spectral method  to
produce power spectra \citep{komm99, sct07}. The mode frequencies and other
mode parameters are estimated by fitting the individual peaks \citep{an90}; not
all modes are fitted successfully at every epoch due to the stochastic
excitation nature
of the modes. In this paper, we concentrate on the behavior of individual modes
rather than the average quantities. Although each data set has a large
number of  multiplets, only 19 multiplets are found to be common to all of the
data sets and have an inner turning point half-way between the tachocline and
solar surface.  These multiplets cover a frequency ($\nu$) range of 2138 --
3362~$\mu$Hz and degree ($\ell$) range between 50 and 81. 

The mode parameters are correlated with four well known surface activity
indicators: the
integrated radio flux at 10.7 cm ($F_{10}$) obtained from Solar Geophysical
Data (SGD), Magnetic Plage Strength Index (MPSI) from Mt. Wilson  magnetograms
\citep{ul91}, the International Sunspot Number ($R_I$) obtained from SGD and
Mg {\sc II} core-to-wing ratio \citep{viereck01}
\begin{figure}[!t]
\plottwo{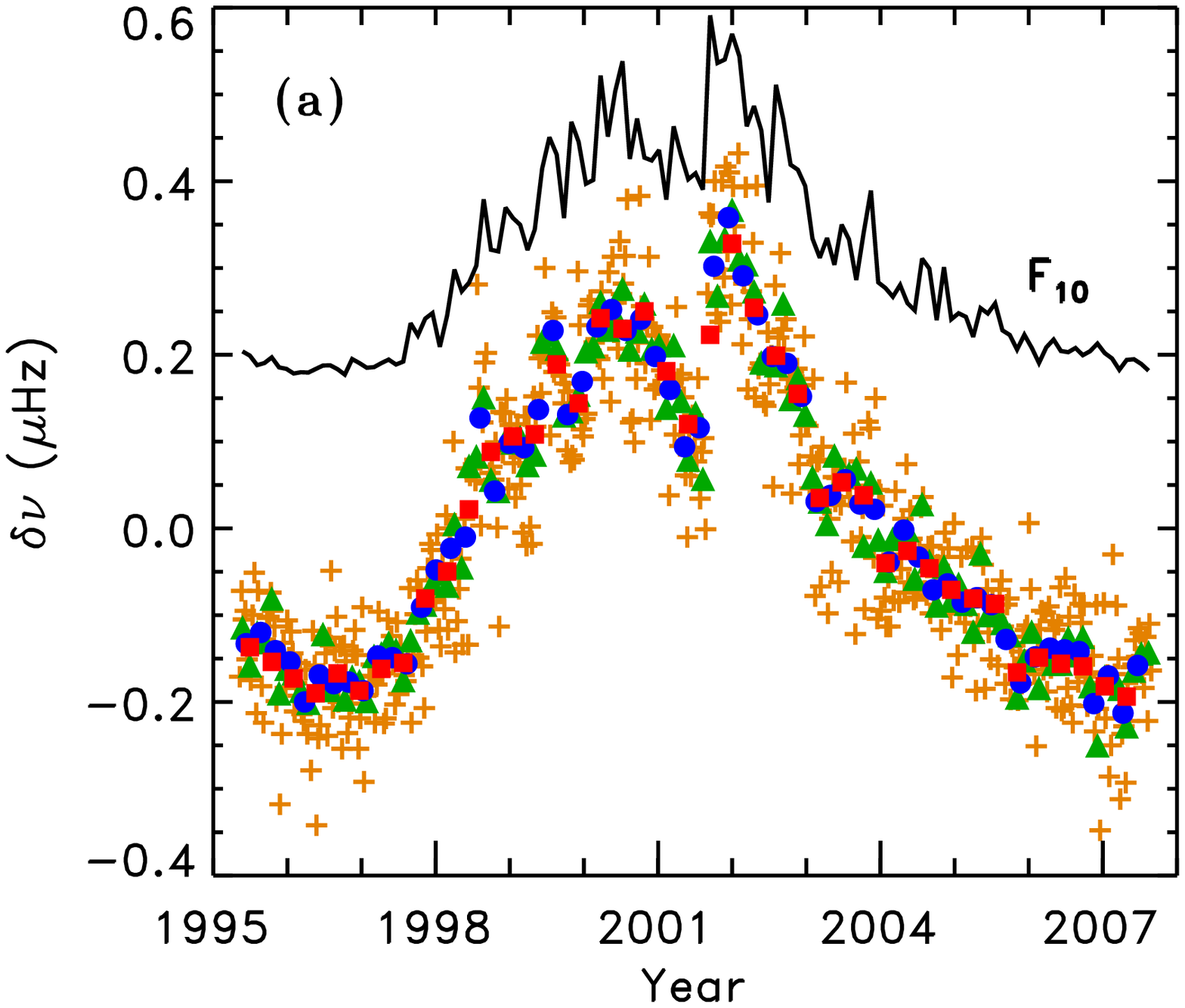}{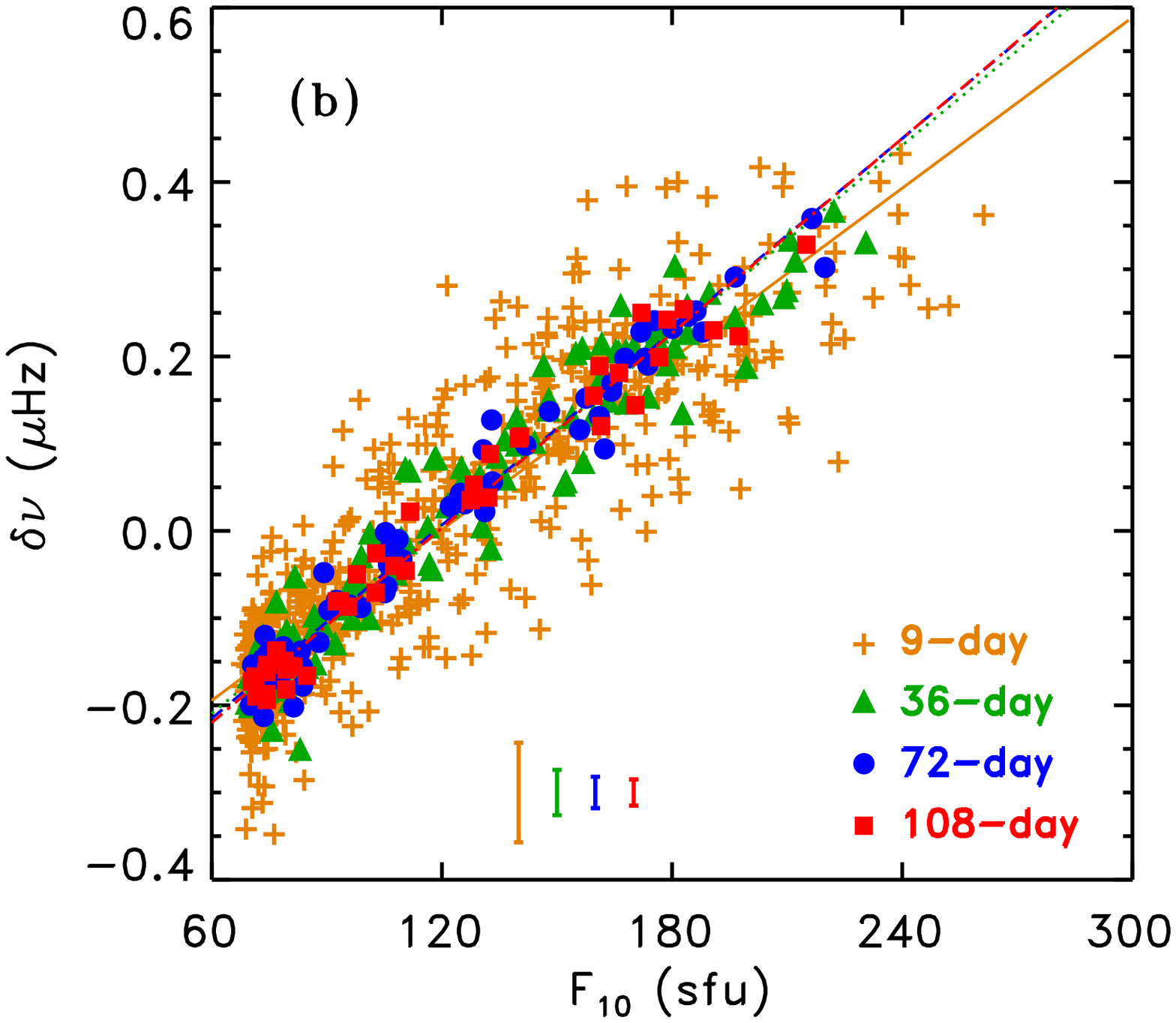}
\caption{(a) The temporal evolution of the  frequency shifts 
corresponding to the multiplet $n$~=~8, $\ell$~=~72.  The solid line 
shows the solar activity. 
(b) The shifts as a function of the radio flux for the same mode. The solid
lines denote the linear fits between them. The vertical lines on the bottom
indicate 1$\sigma$ errors in the increasing order of the
length of the time series. Different symbols indicate different lengths and
are marked in (b). \label{fig1}}
\end{figure}

\section{Results}
Figure~1(a) shows the temporal evolution of frequency shifts 
 corresponding to the $n$~=~8,
$\ell$~=~72 multiplet with respect to the average frequency of the mode;
different symbols correspond to time series of different lengths.
A distinct temporal variation depicting the solar activity cycle is easily
observed in each data sets. Although the modes are well resolved in the 9-day
time
samples, a large scatter is seen for the frequencies,  
probably due to the lower frequency resolution and broader line widths.  
Figure~1(b)  demonstrates the linear relationship between the frequency shifts
and 10.7~cm radio flux measurements. 

As an example of the relation between the frequency shifts and activity
indices, the left panel of Figure~2 shows the correlation  
coefficients for  four different multiplets as a function of the length of the
time series.  For all of the activity indices considered here, the correlation
improves with the length but the nature of the increase is significantly
different for different multiplets. For example the $n$~=~8, $\ell$~=~81
multiplet
shows a steep increase in correlation from 9 to 108 days while for the $n$~=~8,
$\ell$~=~63 
multiplet, the correlation flattens after 36 days.  
The variations of the slope as a function of the correlation
coefficients for different data sets are shown in the right panel of Figure~2.
Each
point in the figure represents a common multiplet. We find that the
correlation and slopes are correlated and both increase from 9 days to
108 days. However, our earlier studies \citep{sct07} involving average
frequency shifts had shown that the slopes obtained from 9 days are higher
than those from 108 days. Thus,  the behavior of individual modes appear to
be different from their average quantities. 

\begin{figure}[!t]
\plottwo{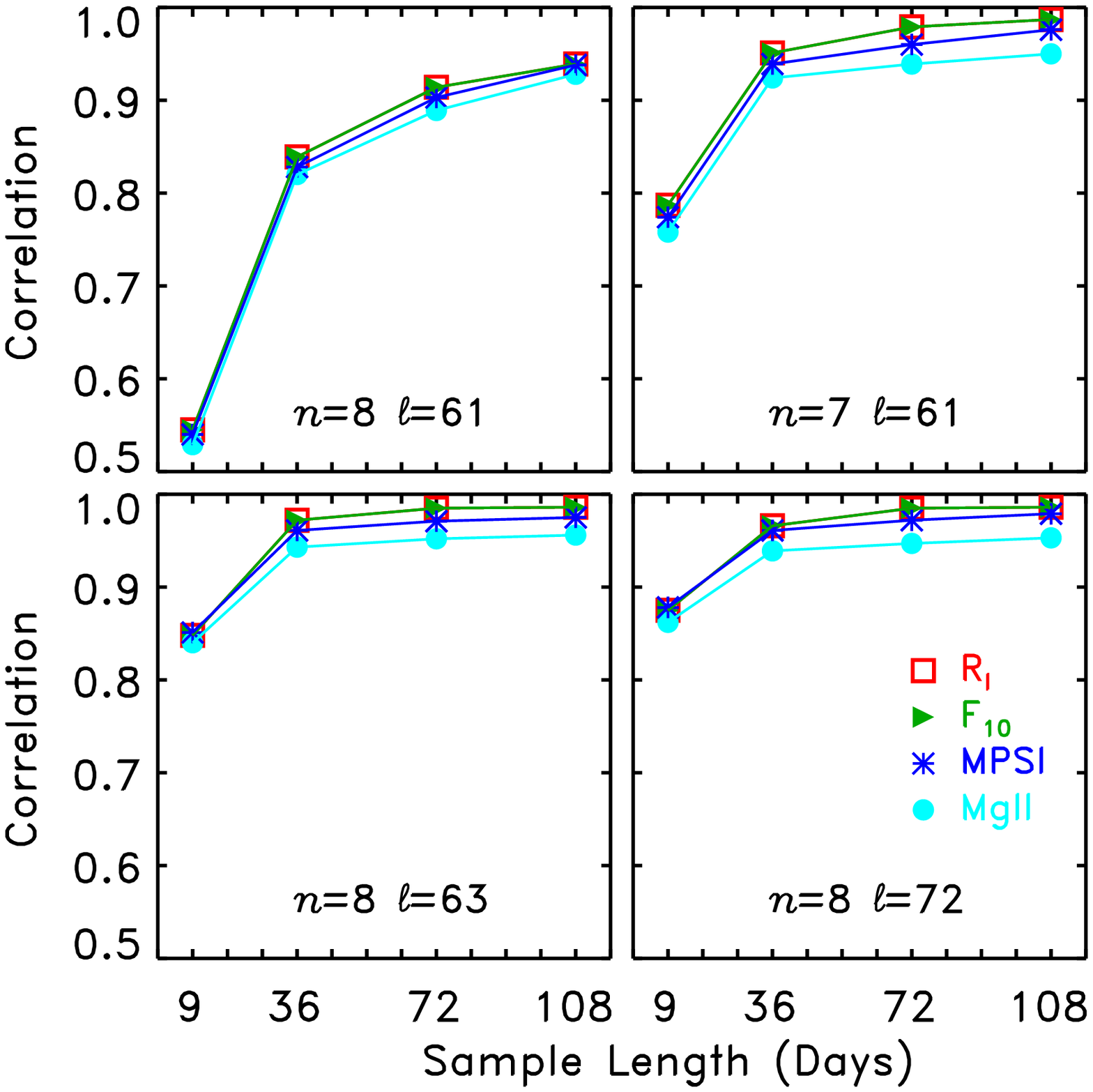}{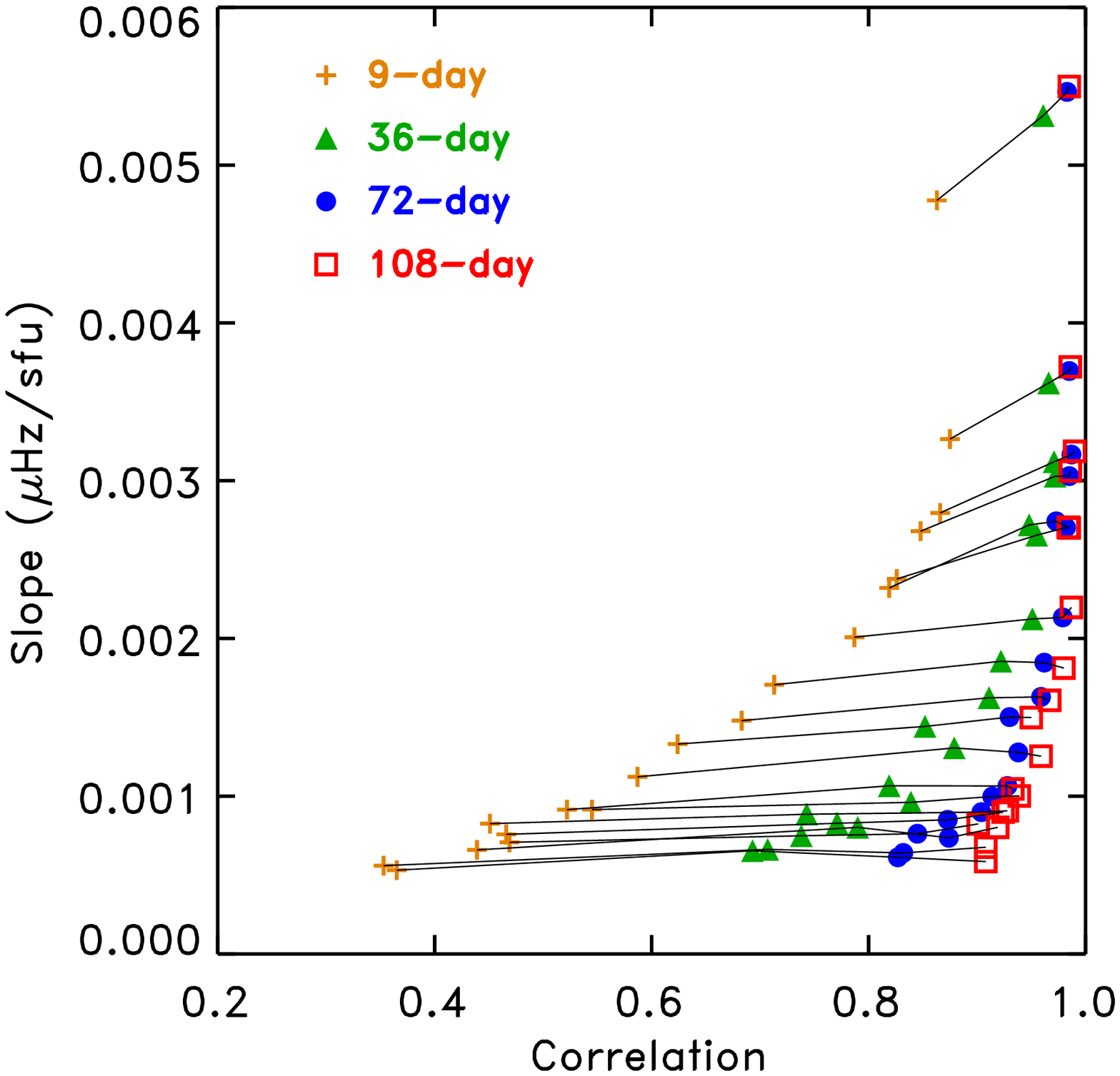}
\caption{The left panels shows the correlation coefficients obtained from the
linear fit between frequency shifts and different activity indices (marked in
the bottom panel) for four different multiplets as a function of sample length.
The right panel shows the
slopes obtained from $F_{10}$ 
as a function of the
correlation coefficient; each point denotes a common multiplet.  The symbols
represent different time segments and are indicated in the panel. \label{fig2}}
\end{figure}

\begin{figure}[!t]
\plottwo{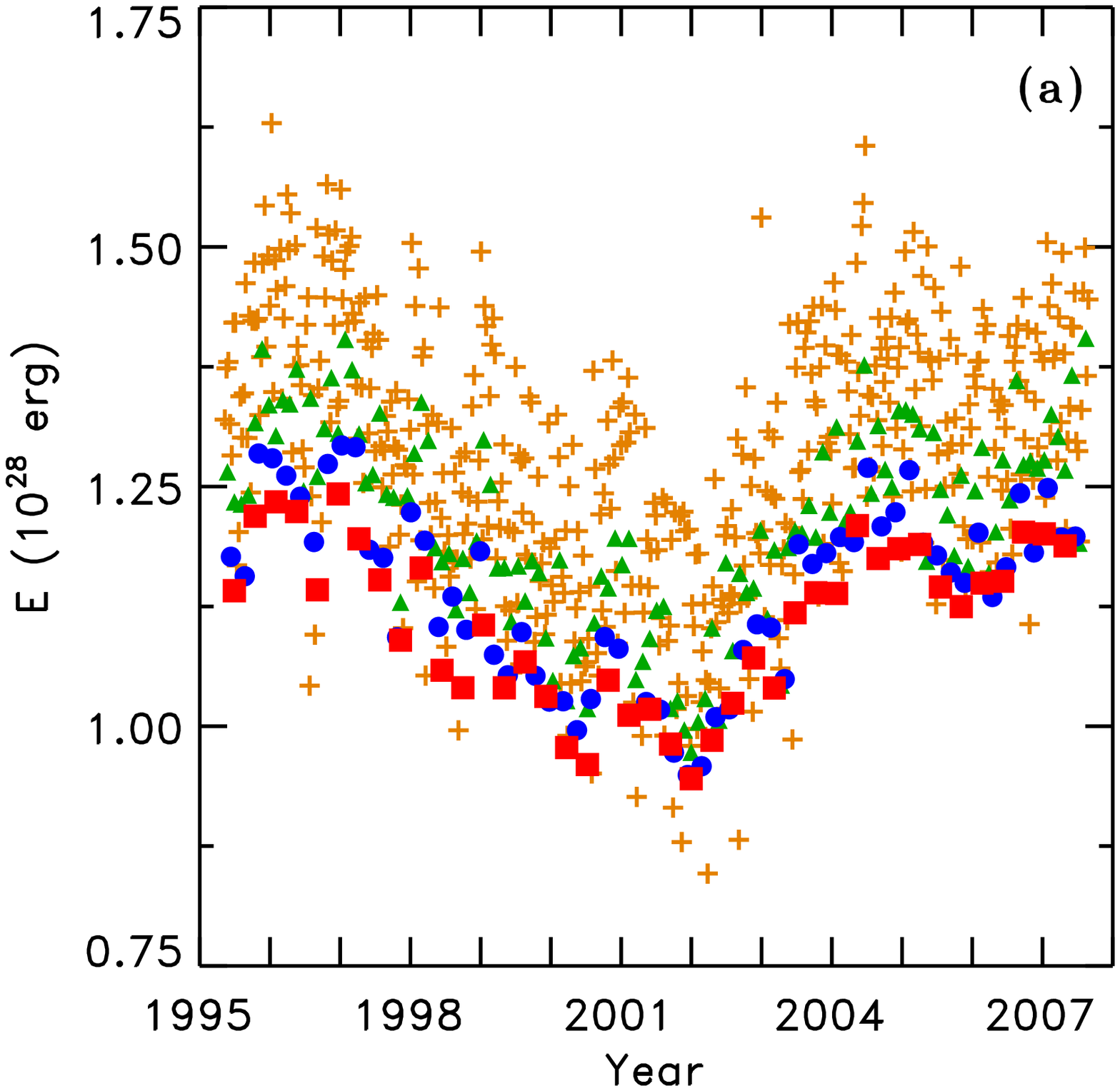}{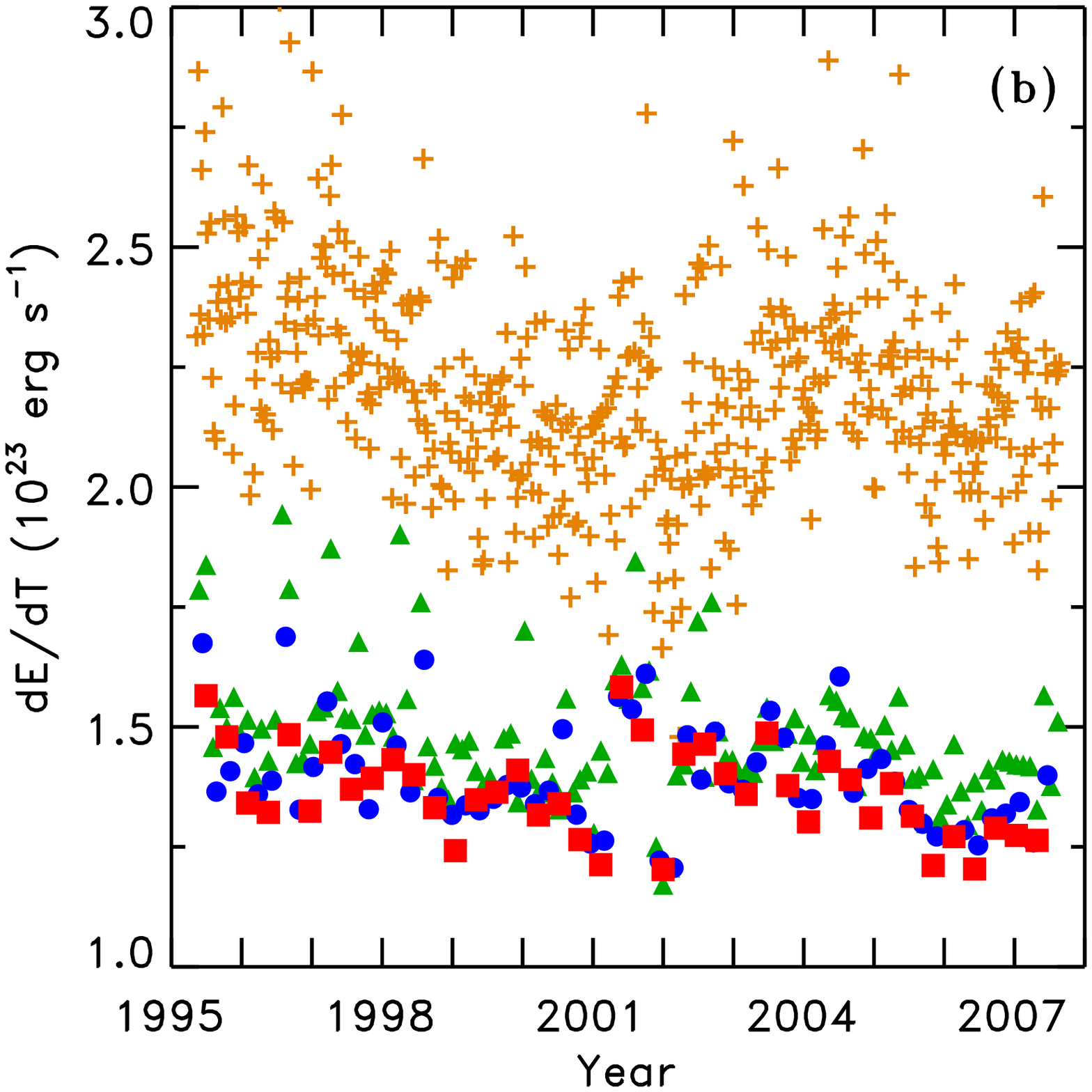}
\caption{The temporal evolution of (a) mode energy and (b) energy supply rate  
for the $n$~=~8, $\ell$~=~72 multiplet. The symbols have the same
meaning as in  Figure~1.  
\label{fig3}}
\end{figure}
We correct the mode amplitudes and line widths for  gaps in the temporal window
function  and then combine them to estimate the mode energy ($ \propto A
\times \Gamma$) and energy supply rate ($ \propto A \times \Gamma^2$) assuming 
stochastic excitation \citep{gold94}. The temporal evolution of these
parameters for the  $n$~=~8, $\ell$~=~72 multiplet is shown in Figure~3. As
discussed in \citet{komm02}, the plots demonstrate an annual
modulation. We further observe that the energy of the mode (Fig.~3a)
is anti-correlated with solar activity with correlation coefficients of
$-$0.62, $-$0.83, $-$0.86, and $-$0.90 with $F_{10}$ for 9 to 108~days
respectively. This again demonstrates that the correlation increases with the
length of the time series. On the other hand, the energy supply rate (Fig.~3b)
shows only short-term variations and no apparent correlation with the
solar activity cycle. These general behaviors are consistent with those
reported by \citet{chap00} for the low-degree modes.  Further, we also note
differences between energy and energy supply rate obtained from time series of
different lengths; the values are higher for shorter time series. Similar to
the mode frequency,  it is also possible that different modes may behave
differently and as a result the average behavior may be different than for 
individual multiplets.   

\section{Conclusion}
Analyzing mode parameters of individual multiplets for data sets of different
lengths, we find that the slope and linear correlation coefficients between
mode frequencies and activity indicators increases with the length of the time
series.   In all cases, the slopes are found to be correlated with the
correlation
coefficients.  We also observe that the mode energy decreases with increase in
solar activity while the energy supply rate is approximately constant over the
entire solar cycle; the values of these parameters are progressively higher
for shorter time series.  The study also indicated that the behavior of
individual modes may be different from their average behavior.  
Additional work is in progress to confirm these findings.

\acknowledgements 
This work utilizes data obtained by the Global Oscillation Network
Group program, managed by the National Solar Observatory, which
is operated by AURA, Inc. under a cooperative agreement with the
National Science Foundation. The data were acquired by instruments
operated by the Big Bear Solar Observatory, High Altitude Observatory,
Learmonth Solar Observatory, Udaipur Solar Observatory, Instituto de
Astrof\'{\i}sica de Canarias, and Cerro Tololo Interamerican
Observatory.  This study includes data from the synoptic program at the
150-Foot Solar Tower of the Mt. Wilson Observatory. The Mt. Wilson 150-Foot
Solar Tower is operated by UCLA, with funding from NASA, ONR, and NSF, under
agreement with the Mt. Wilson Institute. This work is supported in part by
NASA grant  NNG 05HL41I.

\end{document}